# Isotope exponent in disordered underdoped and overdoped La214

**S H Naqib\* and R S Islam**

*Department of Physics, Rajshahi University, Rajshahi-6205, Bangladesh*

\*E-mail: salehnaqib@yahoo.com

**Abstract**

The effect of oxygen isotope substitution on $T_c$ has been analyzed for heavily underdoped and overdoped $La_{2-x}Sr_xCu_{1-y}Zn_yO_4$ compounds with different Zn contents in the $CuO_2$ plane. The effect of Zn on the isotope exponent, $\alpha$, was more pronounced in the case of the underdoped ($x = 0.09$) compounds compared to the overdoped ($x = 0.22$) ones. The variation of $\alpha$ with in-plane disorder can be described quite well within a model based solely on impurity induced Cooper pair-breaking in the case of the underdoped compounds. This model, on the other hand, fails to describe the behavior of $\alpha(y)$ for the overdoped samples, even though Zn still breaks pairs very effectively at this hole (Sr) content. We discuss the implications of these finding in details by considering the Zn induced magnetism, stripe correlations, and possible changes in the superconducting order parameter as number of charge carriers in the $CuO_2$ plane, $p$ ($\equiv x$), is varied.

## 1. Introduction

The physical mechanism responsible for superconductivity in high-$T_c$ cuprates remains a mystery still after a quarter of a century since their discovery. It is widely believed that the normal and superconducting (SC) state properties of these materials are non-Fermi liquid like [1, 2] and the conventional BCS theory [3] fails to describe the various features shown below the SC transition temperature. Investigation of the isotope effect, which played a major role in the development of the BCS theory within the framework of electron-phonon coupling [4], is highly confusing in the case of cuprates [5 – 7]. The possible role played by phonons on the appearance of superconductivity in cuprates is a matter of strong debate [5 – 10]. In these strongly correlated electronic systems, the isotope exponent (IE), $\alpha$, depends largely on the number of added holes in the CuO$_2$ plane, $p$. $\alpha(p)$ increases in the underdoped (UD) side and can significantly exceed the canonical BCS value of 0.5 near the so-called 1/8$^{th}$ anomaly [10, 11]. The value of IE near the optimum doping, on the other hand, is extremely small and stays almost the same (or increases slightly) in the overdoped (OD) region [6, 10]. The problem in interpretation of the IE is complicated because of the various electronic correlations present in cuprates. The IE is affected both by spin/charge ordering (stripe correlations) and pseudogap (PG) in the quasiparticle energy spectrum [10]. How these correlations are related to superconductivity itself is not well-understood yet. Besides, pair-breaking due to impurities and the symmetry of the order parameter also play significant roles [9, 12]. There is renewed interest in possible role of lattice on the quasiparticle dynamics in cuprates following the isotope dependent ARPES results [13]. A clear understanding of the hole content and disorder dependences of the IE is important to clarify various issues regarding carrier pairing and to reveal the possible roles of the underlying electronic correlations present in high-$T_c$ cuprates.

In this study we have analyzed the oxygen ($^{16}$O→ $^{18}$O) isotope effect on the $T_c$ of La$_{2-x}$Sr$_x$Cu$_{1-y}$Zn$_y$O$_4$ (Zn-La214) compounds. Two different Sr contents, $x$ = 0.09 (UD) and $x$ = 0.22 (OD), are used. La214 is a particularly interesting system as strong stripe correlations in both spin and charge channels have been well established [14]. Formation of charge/spin stripes lead to a strong coupling between electronic and lattice degrees of freedom and IE is expected to get enhanced. The UD compounds used in this study has a substantial PG, but PG is absent for the OD compounds. The existence and the extent of the PG can increase the IE appreciably [10]. For the Zn-free La214, we have found the oxygen isotope exponent (OIE) substantially higher in the UD compound compared to the OD one. Zn substitution further increases OIE systematically for both the UD and the OD compounds but the rate of increment is much higher for the UD ones, even though Zn suppresses $T_c$ very effectively in the OD compounds. The variation of $\alpha$ in the UD compounds with disorder can be described well using a model based solely on pair-breaking [9, 12], but this model fails completely to describe the behavior of $\alpha(y)$ for the OD compounds. We discuss the possible implications of these in the subsequent sections.

## 2. Experimental samples

Polycrystalline, single-phase, sintered compounds with almost identical density were used. Results obtained for heavily underdoped $La_{1.91}Sr_{0.09}Cu_{1-y}Zn_yO_4$ (with Zn contents $y$ = 0.00, 0.005, and 0.01) and heavily overdoped $La_{1.78}Sr_{0.22}Cu_{1-y}Zn_yO_4$ (with $y$ = 0.00, 0.01, and 0.02) are reported in this paper. Details of sample preparation, characterization, and isotopic substitution procedure can be found in ref. [15]. Almost identical value of the room-temperature thermopower, $S$[290 K], showed that the $p$-values did not change appreciably with Zn substitution for fixed value of Sr content [16]. $T_c$, on the other hand, was suppressed rapidly by Zn substitution. The rate of suppression was strongly dependent on the hole content, $dT_c/dy \sim$ - 15 K/%Zn and - 8 K/%Zn for the UD and the OD compounds, respectively.

The level of oxygen isotope enrichment was determined from the weight changes of the $^{16}O$ and $^{18}O$ exchanged samples. Raman spectroscopy was also used [15] as a means to double-check if samples were successfully isotope-exchanged, as any isotopic substitution will induce shifts in the associated Raman lines. The levels of isotope substitution found from the weight changes (Raman shifts) were: 99% (98%) for $La_{1.91}Sr_{0.09}CuO_4$; 99% (96%) for $La_{1.91}Sr_{0.09}Cu_{0.995}Zn_{0.005}O_4$; 97% (93%) for $La_{1.91}Sr_{0.09}Cu_{0.99}Zn_{0.01}O_4$ and 100% (96%) for $La_{1.78}Sr_{0.22}CuO_4$; 93% (93%) for $La_{1.78}Sr_{0.22}Cu_{0.99}Zn_{0.01}O_4$; 94% (92%) for $La_{1.78}Sr_{0.22}Cu_{0.98}Zn_{0.02}O_4$. These values are accurate within 5%. To check the reliability of the IE data shown here, we have back-exchanged ($^{18}O \rightarrow {}^{16}O$) two of the randomly selected compounds with different Sr contents and have found highly reproducible results.

## 3. Experimental results

IE was obtained from the DC magnetization measurements using a SQUID magnetometer (*Quantum Design MPMS2*). The field-cooled mode was employed using a DC magnetic field of 150 Oe parallel to the longest sample dimension. The *M-T* curves for the 9% and 22% Sr compounds with different Zn contents are shown in Fig. 1a and 1b, respectively. The back-exchanged data for a representative pair of samples is shown in Fig. 2, showing that the observed isotope effect is not due to the differences in grain size, demagnetization factor and remanent magnetization of the $^{16}O$ and $^{18}O$ sample pairs. $T_c$ is taken at the point of intersection between the line extrapolated from the steepest part of the magnetization curve and the horizontal line of almost zero magnetization in the normal state [8, 17]. An example is shown in the inset of Fig. 1a. The isotope exponent was calculated from the expression $\alpha = - m\Delta T_c/T_c\Delta m$. We show the effects of hole content and in-plane disorder on the IE in Fig. 3. The IE for the pure compound, $\alpha_p$, is almost three times larger (0.271) for the UD one compared to the OD sample (0.0939).

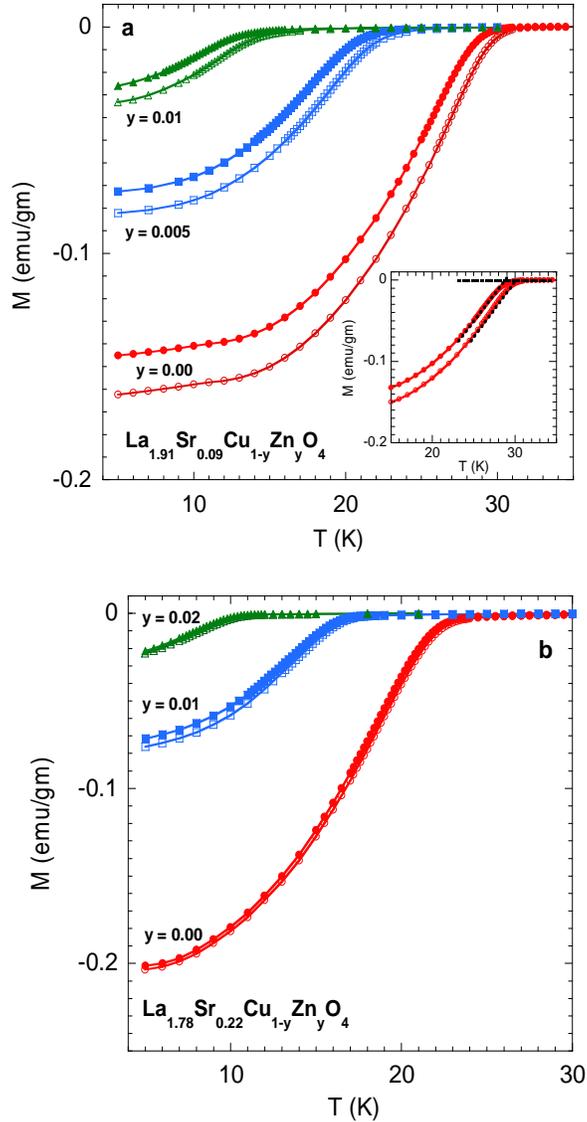

Figure 1 (color online): $M$-$T$ data for (a) UD Zn-La214 and (b) OD Zn-La214. The open symbols are for the samples with $^{16}$O and the filled ones are for the ones with $^{18}$O. The sample compositions are shown in the plots. The inset of (a) shows the method used to determine $T_c$ values (see text for details).

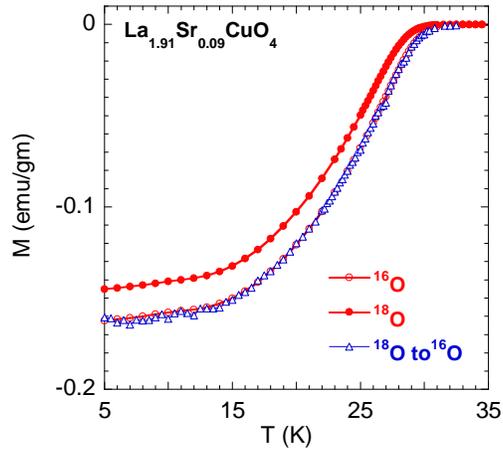

Figure 2 (color online): *M-T* plots for isotope exchanged and back-exchanged pure UD La214.

## 4. Analysis of the IE data

If it is assumed that the same pairing mechanism is at play in the UD and in the OD cuprates (a reasonable assumption), then the main point of concern should be, what are the differences between the UD and the OD samples? And how do those differences affect $T_c$? The major differences between the UD and OD samples are (a) presence of a large PG in the former and its absence in the later [17, 18] and (b) proximity of the UD compound to the $1/8^{th}$ *static* stripe anomaly [14]. There could be other, somewhat less firmly established, electronic band structure related issues and the questions of non-adiabacity and mixed pairing schemes (phonon and non-phonon) that can also affect the values of the IE [9] as hole content varies.

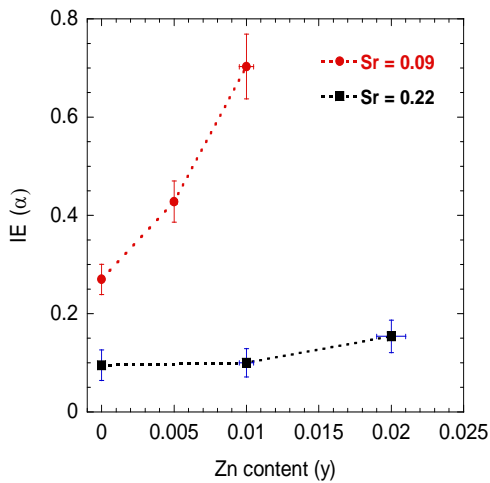

Figure 3 (color online): The IE versus Zn content plot. Sr concentrations are shown. The dashed lines are drawn only as guide to the eyes.

In case the PG reveals itself as a competing correlation, then the spectral weight removed due to its presence will be unavailable to the SC condensate and as a result $T_c$ would be reduced. The extent of this reduction depends on the relative magnitudes of the PG and the SC gap *i.e.* the ratio $E_g/\Delta$, here $E_g$ is the characteristic PG energy and $\Delta$ is the maximum spectral gap at $T = 0K$ [19]. Thus, an isotope effect in $\Delta$ will necessarily generate sizeable isotope effect in $T_c$ and will show a diverging trend as $T_c \rightarrow 0$. Since PG increases with decreasing $p$ ($T_c$ decreases in the UD side), an increase in $\alpha_p$ (the IE of the in-plane disorder free compound) is expected. Williams *et al.* [19] and Pringle *et al.* [20] developed a model where the variation of $\alpha_p$ with $p$ in pure compounds follows from the $p$-dependence of the $E_g$. In this scenario the IE in the SC gap ($\alpha_\Delta$) is assumed to be a constant, around 0.06, independent of the hole content and the observed variation in the IE in $T_c$ with hole content is due to the variation in the PG energy scale. In the heavily OD compounds, where PG is absent, $\alpha_p \equiv \alpha_\Delta$ [19, 20]. The origin and significance of this small *intrinsic p*-independent value of isotope exponent is unclear though.

It is generally agreed upon that static stripe is detrimental to superconductivity and the lattice plays an important role in the stabilization of the stripe order. Therefore, an enhanced IE in the vicinity of the $1/8^{th}$ doping is likely. There are indications that PG and stripe ordering originates from different physical mechanisms [9, 21]. In this situation the contribution to the IE originating from the PG and the stripe correlations should be additive and the significantly larger value (compared to the OD one) of $\alpha_p$ for the UD compound might be qualitatively accounted for.

Several earlier studies [9, 12] have analyzed the variation of the IE in disordered cuprates considering only the role played by pair-breaking due to intrinsic or extrinsic disorder. The general trend being, as $T_c$ is decreased due to impurity scattering, IE increases. It is important to realize that both PG and stripe correlations can be incorporated directly to pair-breaking models. In the unitary limit, PG determines the impurity scattering rate ([22], stripe ordering on the other hand, reduces both $T_c$ and the superfluid density around the $1/8^{th}$ doping [23]. The combined effect of static stripe ordering and in-plane impurity is complex as in-plane disorder can change the strength of stripe correlations [21]. An important consequence of the magnetic pair-breaking model for the IE is that $\alpha$ can be expressed as a universal function $\alpha(T_c)$ [9]. This follows directly from the definition of the isotope exponent and the Abrikosov-Gorko'v equation [24], given by

$$\ln\left(\frac{T_c}{T_{c0}}\right) = \kappa\left[\psi\left(\frac{1}{2}+\gamma\right) - \psi\left(\frac{1}{2}\right)\right] \tag{1}$$

Here $T_{c0}$ is the SC transition temperature of the pure compound and $\kappa$ ($\leq 1$) is a constant depending on the symmetry of the SC order parameter and the nature of the pair-breaking mechanism. $\gamma = \Gamma/2\pi T_c$, where $\Gamma$ is the pair-breaking scattering rate proportional to the impurity concentration and $\psi$ is the *di-gamma* function. Consequently, one gets [9]

$$\frac{\alpha}{\alpha_p} = \frac{1}{1 - \kappa\psi'\left(\frac{1}{2}+\gamma\right)\gamma} \tag{2}$$

$\psi'$ is the first derivative of $\psi$. Eqs. 1 and 2 are also valid for non-magnetic potential scattering for *d*-wave superconductors [25]. Therefore, irrespective of the nature of the pair-breaking impurity scatterer, the IE in cuprates can be expressed in terms of the experimentally measurable parameters $T_c$ and $\alpha_p$. Here $\alpha_p$ denotes the IE of the pure (*e.g.*, Zn free, in this study) compound. This universal dependence of $\alpha(T_c)$ is shown in Fig. 4. In the same plot, the IE data for all the $La_{2-x}Sr_xCu_{1-y}Zn_yO_4$ compounds are exhibited. $\alpha(y)$ values for the UD compounds agree remarkably well with the universal curve whereas $\alpha(y)$ for the OD compounds falls significantly below this trend.

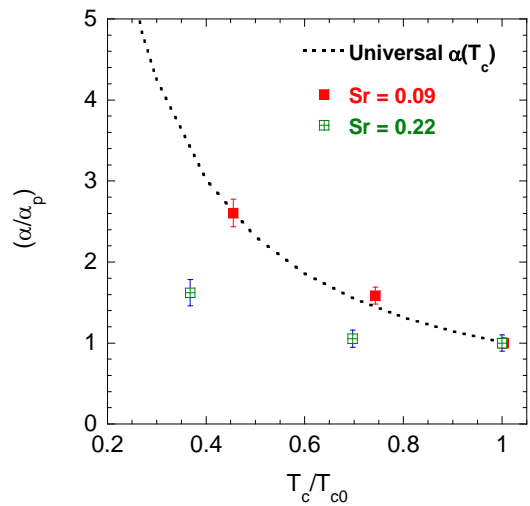

Figure 4 (color online): $\alpha/\alpha_p$ versus $T_c/T_{c0}$ for $La_{1.91}Sr_{0.09}Cu_{1-y}Zn_yO_4$ and $La_{1.78}Sr_{0.22}Cu_{1-y}Zn_yO_4$. The dashed curve shows the universal $\alpha(T_c)$ based on pair-breaking model (using Eq. 2, with $\kappa = 1$).

## 5. Discussion and conclusions

The results shown in Fig. 4 are surprising in several respects. Compared to the UD compounds, the physics of the OD ones are largely free from possible complications due to a sizeable PG and the relevance to the stripe ordering. Zn induced enhancement of magnetic susceptibility found in the UD compounds is also greatly reduced in the OD compounds [26]. As in the case for the UD region, the rate of suppression of $T_c$ with Zn in the OD side can be modeled equally well within the framework of unitarity scattering in $d$-wave superconductors [22]. Within this context there is apparently no reason why this simple pair-breaking model would fail to describe the $\alpha(y)$ behavior for the OD compounds. This discrepancy is the central result of the present paper.

It is interesting to note the dependence of $\alpha/\alpha_p$ on the value of $\kappa$ (Eq.2). In $s$-wave superconductors with non-magnetic impurities, $\kappa = 0$, as a result there is no enhancement in the IE due to pair-breaking. For anisotropic $s$-wave or mixed (e.g., $dx^2$-$y^2$ + $s$) type order parameter, the value of $\kappa < 1$ and the enhancement of $\alpha/\alpha_p$ could be significantly lower with increasing impurity concentration. In fact such a hole content dependent change in the SC order parameter was found in OD Y(Ca)123 [27], where it was found that the SC order parameter is $dx^2$-$y^2$ type for optimally and UD Y(Ca)123 but it consists of a significant $s$ wave component as one moves into the OD region (orthorhombic structure). Therefore, for OD Y(Ca)123 it would not be surprising if $\alpha/\alpha_p$ exhibits a weaker impurity content (or $T_c/T_{c0}$) dependence. Unfortunately this line of reasoning does not help in understanding the behavior of the IE in the UD and OD La214. Unlike in Y123, the OD La214 becomes tetragonal. The low-$T$ orthogonal phase does not exist for $x > 0.20$ [28]. The failure of pair-braking model to describe the IE in disordered (Ni and Fe doped) moderately OD ($x = 0.20$) La214 has been reported before [29]. In that study [29] the authors tried to describe the IE of Fe substituted La214 using the stripe scenario. This type of analysis is questionable since stripe correlations are not expected to play a large role in the OD compound. Also, Fe can change the in-plane hole content in addition to contributing in pair-breaking. Surprisingly the change in the IE with Ni doping was found to be minimal in the OD La214, and in fact $\alpha/\alpha_p$ decreased gradually with increasing Ni content [29, 30]. Effect of Zn on the optimally doped (OPD, $x = 0.15$) and moderately OD La214 was also studied earlier in ref. [30] where the authors tried the explain the somewhat reduced (compared to the OPD ones) values of $\alpha/\alpha_p$ for the moderately OD compounds by invoking to a peak in the EDOS for the OPD compounds. Such an analysis is also debatable because the ARPES results [31] showed that the Fermi level crosses the peak in the EDOS at $p \sim 0.225$.

In summary, the present work reports the IE of UD and OD La214 with different amount of in-plane iso-valent disorder. The evolution of $\alpha(y)$ for the UD compounds can be described well within the pair-breaking model for $d$-wave superconductors. The gradual increment of the $\alpha(y)$ for the OD compounds, on the other hand, does not agree with the pair-breaking calculations, even though significant pair-breaking is present in these samples due to Zn. We have taken into account of the various other mechanisms that can affect the IE in cuprates, but none of these schemes provide with a satisfactory explanation.

**Acknowledgements**

The authors acknowledge the Commonwealth Commission, UK, Trinity College, University of Cambridge, UK, the Industrial Research Limited and MacDiarmid Institute for Advanced Materials and Nanotechnology, Wellington, New Zealand, for financial support and for the experimental facilities.